\documentclass[12pt,preprint]{aastex}
\usepackage{amssymb,amsmath}
\usepackage{color}
\usepackage{hyperref}
\definecolor{linkcolor}{rgb}{0,0,0.25}
\hypersetup{
  colorlinks=true,        
  linkcolor=linkcolor,    
  citecolor=linkcolor,    
  filecolor=linkcolor,    
  urlcolor=linkcolor      
}
\newcommand{\ie}{i.e.}
\newcommand{\etal}{et al.}
\newcommand{\eg}{e.g.}
\newcommand{\eqnname}{equation}

\newcommand{\sectionname}{$\mathsection$}
\newcommand{\tvector}[1]{\boldsymbol{\vec{#1}}}
\newcommand{\vx}{\tvector{x}}
\newcommand{\vv}{\tvector{v}}
\newcommand{\vxi}{\tvector{x}_i}
\newcommand{\vvi}{\tvector{v}_i}

\newcommand{\setofxv}{\{\vx_i,\vv_i\}}

\newcommand{\lsr}{LSR}
\newcommand{\vsunlsr}{\ensuremath{v_\odot}}

\newcommand{\Ro}{\ensuremath{R_0}}

\renewcommand{\vec}[1]{\mathbf{#1}} 

\newcommand{\masermean}{\ensuremath{\overline{\vv}}}

\newcommand{\vvpec}{\ensuremath{\vv_{\mbox{{\footnotesize pec}}}}}

\newcommand{\vvmaser}{\ensuremath{\vv_{\mbox{{\footnotesize maser}}}}}
\newcommand{\eeR}{\ensuremath{\vec{e}_{R}}}
\newcommand{\eephi}{\ensuremath{\vec{e}_{\phi}}}
\newcommand{\eez}{\ensuremath{\vec{e}_{z}}}

\newcommand{\maserdisp}{\ensuremath{\mbox{\boldmath$\sigma$}}}

\newcommand{\normal}{\ensuremath{\mathcal{N}}}

\newcommand{\trace}{\mbox{Trace}}

\newcommand{\pmsgra}{\ensuremath{\mu_{\mbox{{\footnotesize Sgr A$^*$}}}}}

\newcommand{\vlba}{VLBA}
\newcommand{\vlbi}{VLBI}
\newcommand{\vera}{VERA}
\newcommand{\reid}{R09}
\newcommand{\vc}{V_c}

\newcommand{\ra}{\ensuremath{\alpha}}
\newcommand{\dec}{\ensuremath{\delta}}

\newcommand{\degree}{^{\circ}}

\begin{document}

\title{Galactic masers and the Milky Way circular velocity}
\author{Jo~Bovy\altaffilmark{1,2},
  David~W.~Hogg\altaffilmark{1,3},
  Hans-Walter Rix\altaffilmark{3}
}
\altaffiltext{1}{Center for Cosmology and Particle Physics, Department of Physics,
  New York University, 4 Washington Place, New York, NY 10003, USA}
\altaffiltext{2}{Correspondence should be addressed to jo.bovy@nyu.edu~.}
\altaffiltext{3}{Max-Planck-Institut f\"ur Astronomie, K\"onigstuhl 17, D-69117 Heidelberg, Germany}

\begin{abstract}
Masers found in massive star-forming regions can be located precisely
in six-dimensional phase space and therefore serve as a tool for
studying Milky Way dynamics. The non-random orbital phases at which
the masers are found and the sparseness of current samples require
modeling. Here we model the phase-space distribution function of 18
precisely measured Galactic masers, permitting a mean velocity offset
and a general velocity dispersion tensor relative to their local
standards of rest, and accounting for different pieces of prior
information. With priors only on the Sun's distance from the Galactic
Center and on its motion with respect to the local standard of rest,
the maser data provide a weak constraint on the circular velocity at
the Sun of $\vc = 246 \pm 30$ km s$^{-1}$. Including prior information
on the proper motion of Sgr A$^*$ leads to $\vc = 244 \pm 13$ km
s$^{-1}$.  We do not confirm the value of $\vc \approx 254$ km
s$^{-1}$ found in more restrictive models.  This analysis shows that
there is no conflict between recent determinations of $\vc$ from
Galactic Center analyses, orbital fitting of the GD-1 stellar stream,
and the kinematics of Galactic masers; a combined estimate is $\vc =
236 \pm 11$ km s$^{-1}$. Apart from the dynamical parameters, we find
that masers tend to occur at post-apocenter, circular-velocity-lagging
phases of their orbits.
\end{abstract}


\section{Introduction}

The value of the circular orbital velocity at the Sun's radius in the
Milky Way is of considerable interest in Galactic and extragalactic
astrophysics. It is necessary to correct observed velocities of stars
and galaxies for the motion of the Sun around the Galactic Center. The
circular velocity also plays a large role in characterizing the mass
of the Milky Way in comparison with other spiral galaxies, placing it
in a cosmological context, \eg, when asking whether the Milky Way
matches the Tully--Fisher relation \citep[\eg][]{Klypin02a,Flynn06a} or
what is its total star formation efficiency
\citep[\eg,][]{Smith07a,Xue08a}.

The circular velocity at the Sun's radius has typically been
established by measuring the Sun's motion with respect to an object
assumed to be at rest with respect to the Galaxy (Sgr A$^*$:
\citealt{Reid04a}; the stellar halo: \citealt{Sirko04a}), or by using
a tracer population assumed to be angle-mixed, \ie, having a uniform
distribution of orbital phases, in a steady-state Galaxy
\citep[\eg,][]{Feast97a}. Recently, a competitive estimate has been
obtained by a different approach using a narrow stellar stream that is
assumed to be tracing out an orbit \citep{Koposov09a}.

In this paper we re-analyze a new population of tracers of Milky Way
dynamics: masers associated with star-forming regions
\citep[\reid]{Reid09a}. Using the Very Long Baseline Array (\vlba) and
the Japanese \vlbi\ Exploration of Radio Astronomy (\vera), precise
measurements of the parallaxes, proper motions, and line-of-sight
velocities of masers have been made (see \reid\ and references
therein). These give accurate full six-dimensional phase-space
information in the disk of the Galaxy. Since these massive
star-forming regions are associated with spiral arms and their shocks,
the dense molecular gas regions that produce masers do not lie on
exactly circular orbits, nor are they detected at random points on
their orbits. Therefore, modeling approaches that assume a uniform
distribution of the orbital phases of the tracer population cannot
give accurate determinations of the dynamics of the Galaxy. For the
existing maser data, the problem of non-random orbital phases is
exacerbated by the sparseness of the sample---only 18 masers with
accurate six-dimensional phase-space information have been measured at
present---and by the spatially non-uniform selection of the current
sample of masers.

In this paper, we perform an analysis of the \reid\ maser data that
deals simultaneously with the sparseness of the data, the spatial
non-uniformity of the sampling, the non-random orbital phase
distribution of masers, and prior information. Assuming a flat
rotation curve, $\vc(R) = $ constant, we use a simple model for the
distribution of the maser velocities with respect to their local
standards of rest: a mean offset from circular rotation $\vc(R)$ and a
general velocity dispersion tensor fixed in Galactocentric cylindrical
coordinates. In the probabilistic inference framework that we
use---described in \sectionname~\ref{sec:data}---we can marginalize
over the uncertainty in the inferred distribution function of masers,
take prior information on the dynamics of the Galaxy into account, use
the sparse data set as efficiently as possible, and then ask what
information on $\vc$ the maser data provide. Our results presented in
\sectionname~\ref{sec:results} show that allowing for a finite
velocity dispersion tensor in the model for the maser
peculiar-velocity distribution function leads to lower values of $\vc$
than the large value reported in \reid, in whose analysis the maser
velocity dispersion was (implicitly) assumed to vanish.  Adding in
informative prior information about $R_0$, inferred from monitoring
stellar orbits around the black hole at the center of the Galaxy
\citep{Ghez08a,Gillessen09a} and from the measurement of the proper
motion of Sgr A$^*$ \citep{Reid04a}, we find that the best circular
velocity estimate is $\vc = 244 \pm 13$ km s$^{-1}$, but that the
current maser data set adds little information. We discuss this
measurement and its limitations in the light of other recent
determinations in \sectionname~\ref{sec:discussion}.

\section{Data and methodology}\label{sec:data}

Throughout the analysis that follows we use the standard cylindrical
Galactocentric coordinate frame $(R,\phi,z)$, with associated unit
vectors ($\eeR,\eephi,\eez$) pointing toward the Galactic center, in
the direction of Galactic rotation, and toward the North Galactic
Pole, respectively.

\subsection{Data from Reid et al.~(2009)}\label{sec:datasub}

The data we analyze here consist of the Galactic coordinates,
parallaxes, proper motions, and line-of-sight velocities of 18
Galactic masers, as well as their associated uncertainties, presented
in Table 1 of \citet{Reid09a}. Following \reid, we add a 7 km s$^{-1}$
uncertainty in quadrature to the uncertainties in the velocity
components of each maser to describe the random, virial motion in the
massive star-forming region of the individual massive star associated
with each maser.

The line-of-sight velocities have been `corrected' by the radio
observatories' pipelines for the motion of the Sun with respect to the
Local Standard of Rest (\lsr). This correction assumed a value of 20
km s$^{-1}$ toward \ra(B1900.0)= 18$^\mathrm{h}$,
\dec(B1900.0)=$+30\degree$ for the Solar motion \vsunlsr, although it
is unclear whether all observatories used this standard value
(M.~Reid, private communication). We undo this correction, after
which the currently accepted correction for \vsunlsr\ can be applied;
however, as we will describe below, this correction will become part
of our model and, therefore, the correction for \vsunlsr\ does not
occur during the preprocessing of the data.

Beyond these two corrections, no processing of the \citet{Reid09a}
data has been done.

\subsection{Probabilistic framework}

Parameter estimation in a probabilistic framework \emph{by necessity}
uses Bayes's theorem to connect the probability of the model
parameters given the data
$\{\vxi^{\mathrm{obs}},\vvi^{\mathrm{obs}}\}$ to the probability of
the observed data given the model parameters
\citep[\eg,][]{jaynes}. This requires us (1) to identify all the
parameters that need to be included in the model, (2) to write down
the likelihood of the model and (3) to specify suitable priors for the
model parameters. Although the model space needs to be exhaustive, the
probabilistic framework allows integration over uninteresting
parameters.

Here we put forward a model for the maser kinematics in which the
maser velocities are most easily modeled in Galactocentric cylindrical
coordinates. In order to go from the raw data described in
\sectionname~\ref{sec:datasub} to the velocity of each maser in
Galactocentric coordinates, we need to (1) correct the measured
velocity for \vsunlsr, (2) add to this velocity the circular velocity
around the Galactic center at the Sun's radius, and (3) project this
velocity onto the Galactocentric coordinate frame (the details of this
transformation are described in the Appendix of \reid). Since the
latter procedure includes geometrical projection factors depending on
the distance \Ro\ of the Sun from the Galactic Center, the model
parameters need to include the three components of \vsunlsr, \Ro, and
$\vc$. However, it is more practical to assume that Sgr A$^*$ is at
rest with respect to the Galaxy, and to use the proper motion
$\pmsgra$ of Sgr A$^*$ \citep{Reid04a} as a model parameter instead of
the circular velocity, as $\pmsgra$ is very tightly constrained
independently of $R_0$. These two parameters are related simply by
multiplying the proper motion of Sgr A$^*$ by $R_0$ and correcting
this for \vsunlsr. The circular velocity then becomes a parameter
derived from the actual model parameters, which is no problem in the
probabilistic framework, where it is easy to propagate uncertainties
correctly.  As we will assume that the rotation curve is flat, no
extra parameters to model the shape of the rotation curve need to be
included in the model.

If we had uniformly sampled the phase space of masers and full prior
knowledge of the phase-space distribution function of massive
star-forming regions, this would uniquely specify the likelihood of
the model, as the probability of the measured position and velocity of
each maser would simply be given by the distribution function of the
masers convolved with the observational uncertainty. However, we have
neither a uniform sample of masers nor much prior information about
the distribution of masers throughout the Galaxy. To account for the
spatial non-uniformity of the sample we will focus on the distribution
of velocities at the actually observed position of the maser, instead
of using the full six-dimensional phase-space distribution function to
evaluate the likelihood. For this distribution we will assume that it
only depends on the peculiar velocity $\vvpec \equiv \vvmaser - \vc
\cdot \eephi$ of the maser in Galactocentric cylindrical
coordinates. We will assume that this distribution of peculiar
velocities is given by a Gaussian distribution characterized by a
mean, a 3-vector $\masermean$, the offset from circular motion, and a
general velocity dispersion tensor, a symmetric $3 \times 3$ tensor
$\maserdisp$ with six free parameters. Since there have been no
measurements of either the mean offset from circular motion of the
masers or their velocity dispersion, we will use flat priors on these
quantities. This model is essentially a generalization of the model
used in \citet{Reid09a} where the velocity dispersion tensor was
assumed to vanish; this was a poor assumption as we will show below.

The probability of a single maser is thus given by
\begin{equation}\label{eq:onelike}
p(\vxi^{\mathrm{obs}},\vvi^{\mathrm{obs}}|\pmsgra,R_0,\vsunlsr,\masermean,\maserdisp) =
\normal\left(\vvpec[\vx,\vv]|\masermean,\maserdisp\right)\otimes p(\vx,\vv|\vxi^{\mathrm{obs}},\vvi^{\mathrm{obs}})\,,
\end{equation}
where we have suppressed the dependence of $\vvpec$ on the dynamical
parameters, and where the convolution with the observational
uncertainty distribution
$p(\vx,\vv|\vxi^{\mathrm{obs}},\vvi^{\mathrm{obs}})$ has been
included. The posterior distribution for the 14 model parameters is
then given by
\begin{equation}\label{eq:posterior}
p(\pmsgra,R_0,\vsunlsr,\masermean,\maserdisp|\{\vxi^{\mathrm{obs}},\vvi^{\mathrm{obs}}\})
\propto p(\pmsgra,R_0,\vsunlsr)\,\prod_i
p(\vxi^{\mathrm{obs}},\vvi^{\mathrm{obs}}|\pmsgra,R_0,\vsunlsr,\masermean,\maserdisp)\,,
\end{equation}
where the first factor on the right-hand side is the prior probability
distribution for these parameters and the product is the
likelihood. We have used flat priors for $\masermean$ and
$\maserdisp$, which is why they do not appear explicitly.

\begin{figure}
\includegraphics[width=.46\textwidth]{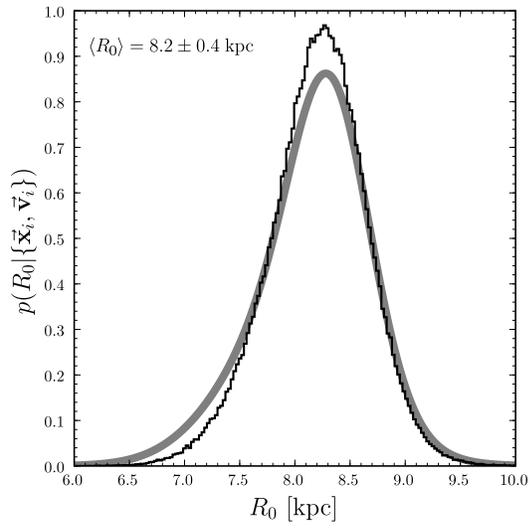}
\caption{Marginalized posterior probability distribution for the
distance \Ro\ to the Galactic center, shown as the black curve, from
10$^6$ MCMC samples. The prior probability distribution is shown as
the thick gray curve; its mean is $R_0 = 8.2 \pm 0.5$
kpc.}\label{fig:ro}
\end{figure}

For $\pmsgra$ we use a Gaussian prior with a mean of 30.24 km s$^{-1}$
kpc $^{-1}$ and a standard deviation of 0.12 km s$^{-1}$ kpc$^{-1}$
\citep{Reid04a}. For \Ro\ we combine current state-of-the-art
determinations of \Ro\ from Galactic Center orbits with equal weights:
8.0 $\pm$ 0.6 kpc found by \citet{Ghez08a} and 8.33 $\pm$ 0.35 kpc
found by \citet{Gillessen09a}. This prior is shown as the gray curve
in \figurename~\ref{fig:ro}. For \vsunlsr\ we use the value and
uncertainties obtained from \emph{Hipparcos} data \citep{Hogg05a},
although the clumpiness of the velocity distribution of nearby stars
\citep{Dehnen98a,Bovy09a} implies an uncertainty more on the order of
a few km s$^{-1}$ in the value of \vsunlsr\ (J.~Bovy \& D.~W.~Hogg, in
preparation). The implied prior for the circular velocity is shown as
the thick gray curve in \figurename~\ref{fig:thetao}. To investigate
how informative the maser measurements are about $\vc$ and $R_0$, we
will consider the effect of dropping (some combination of) these
priors below.

The framework described here can easily be generalized to more general
descriptions of the distribution of the peculiar velocities of the
masers. In what follows we will use a distribution function that is
the sum of two Gaussian distributions, the second having half of the
weight and twice the dispersion of the first Gaussian, to determine
the possible effect of outliers.

\subsection{Exploration of the posterior probability distribution}

In order to explore the posterior distribution for all of the model
parameters in light of the maser data we use a simple Markov Chain
Monte Carlo (MCMC) method \citep{mackay}. This procedure is described
in some detail in the Appendix.

The practical complication in evaluating the likelihood given in
\eqnname s~(\ref{eq:onelike}) and (\ref{eq:posterior}) for each of the
masers comes from the fact that the observational uncertainties are
Gaussian in the space of observed quantities---more specifically, for
the parallax---but are non-Gaussian in the space of the peculiar
velocities. However, if the relative parallax uncertainty is small
($\leq 10$\,percent) we can confidently propagate the uncertainties to
the space of peculiar velocities, where the convolution of the
Gaussian velocity distribution model for the peculiar velocities with
the observational Gaussian uncertainty distribution is simple. A few
of the masers have relative parallax uncertainties larger than
10\,percent, but we have nonetheless propagated the uncertainties in
the Gaussian approximation. To check that this does not bias our
results we have also run our analysis using a full numerical
convolution with the actual observational uncertainties and we find
results that are barely distinguishable from the results presented
below.

\section{Results}\label{sec:results}

\begin{figure}
\includegraphics[width=.46\textwidth]{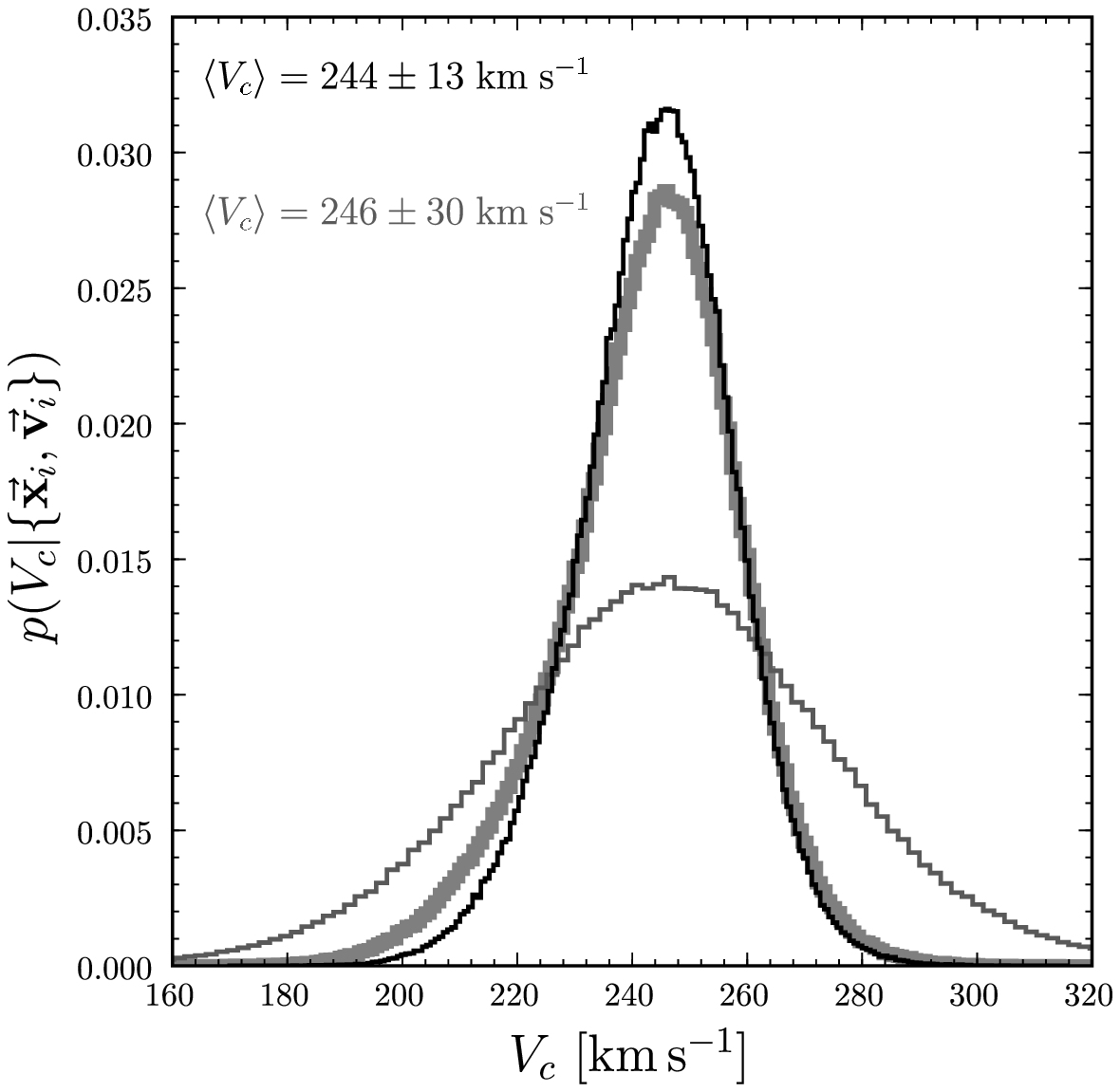}
\caption{Marginalized posterior probability distribution for the
circular velocity $\vc$, shown as the black curve, and its mean (top
label) from 10$^6$ MCMC samples. The prior probability distribution is
shown as the thick gray curve; its mean is $\vc = 243 \pm 16 $ km
s$^{-1}$. The posterior and its mean (bottom label) obtained from
dropping the informative prior on $\pmsgra$ is shown as the thin gray
curve, illustrating that the maser data themselves constrain $\vc$
relatively weakly. The quoted uncertainty in mean value is the
standard deviation $\equiv \sqrt{\langle \vc^2\rangle - \langle
\vc\rangle^2}$.}\label{fig:thetao}
\end{figure}

The main scientific goal of this paper is to understand what the maser
measurements tell us about $\vc$. The posterior probability
distribution for $\vc$, fully marginalized over all of the parameters
of the maser distribution function, the Solar motion with respect to
the \lsr, the distance to the Galactic Center, and the proper motion
of Sgr A$^*$, is shown in \figurename~\ref{fig:thetao}. The
analogously marginalized posterior distribution for \Ro\ is shown in
\figurename~\ref{fig:ro}. Also shown in \figurename~\ref{fig:thetao}
is the posterior we obtained when we drop the informative prior on
$\pmsgra$. The posterior distributions for the proper motion of Sgr
A$^*$ and for the components of \vsunlsr\ are not shown here. They are
all basically identical to their prior distributions, implying that
the masers---not surprisingly---cannot inform us about these
quantities.

While the prior on $\vc$ in \figurename~\ref{fig:thetao} peaks at 244
km s$^{-1}$ with a 1--sigma uncertainty of 16 km s$^{-1}$, the
posterior for $\vc$ is peaked at a value of 244 km s$^{-1}$ with a
1--sigma uncertainty of about 13 km s$^{-1}$. This equal value for
$\vc$ after analyzing the masers is in qualitative contrast to the
initial analysis of \reid, who found that it raised the peak to 254 km
s$^{-1}$. This difference arises mainly from our more general model
for the distribution function of the masers. If we insist within our
analysis that the velocity dispersion of the masers is zero, we find a
posterior distribution for the circular velocity that is peaked at 255
km s$^{-1}$, in rough agreement with the \reid\ results. The light
gray line in \figurename~\ref{fig:thetao} shows what happens when we
drop the informative prior on $\pmsgra$, while keeping the $R_0$
prior: $\vc = 246 \pm 30$ km s$^{-1}$. This and the fact that the
posterior probability is barely narrower than the prior, tells us that
the current maser measurements have not much power to constrain
$\vc$. The posterior estimate for the distance to the Galactic Center
is $R_0 = 8.2 \pm 0.4$ kpc; this shows that the masers lead to a small
improvement to our knowledge of the Sun's distance to the Galactic
Center. Without the informative prior on $\pmsgra$ the posterior
estimate for $R_0$ is the same as the prior estimate: $R_0 = 8.2 \pm
0.5$ kpc.

\begin{figure}
\includegraphics[width=.46\textwidth]{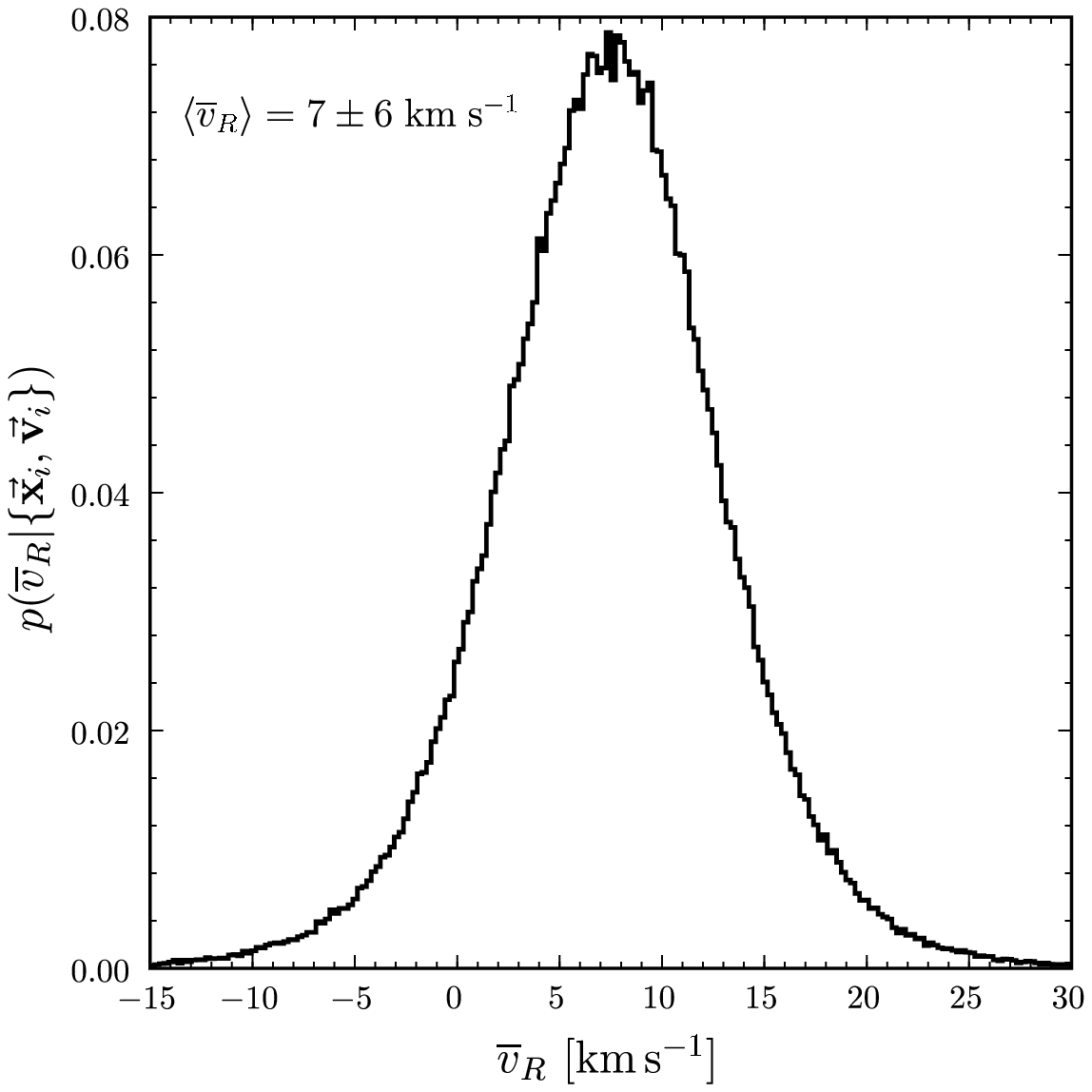}
\includegraphics[width=.46\textwidth]{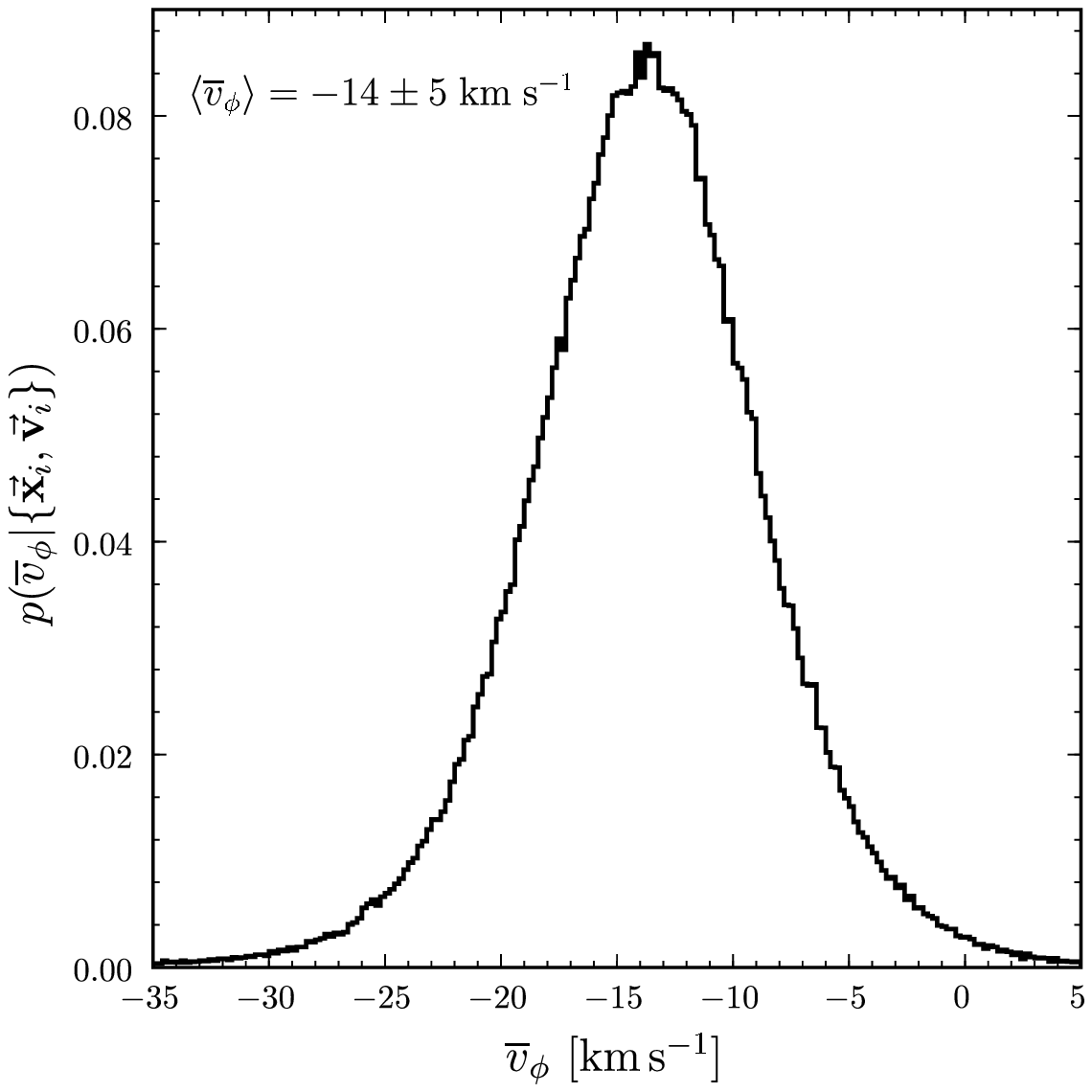}\\
\includegraphics[width=.46\textwidth]{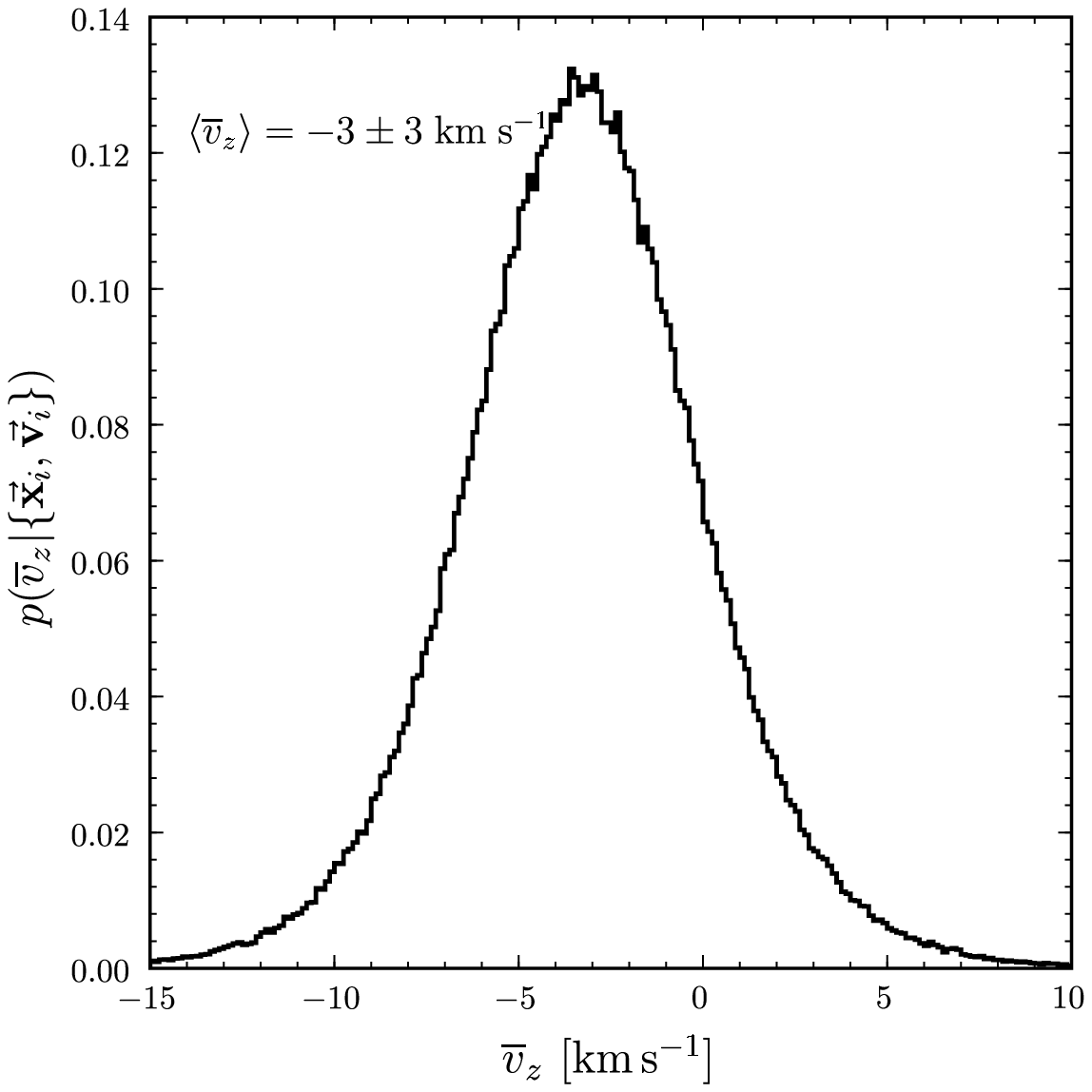}
\includegraphics[width=.46\textwidth]{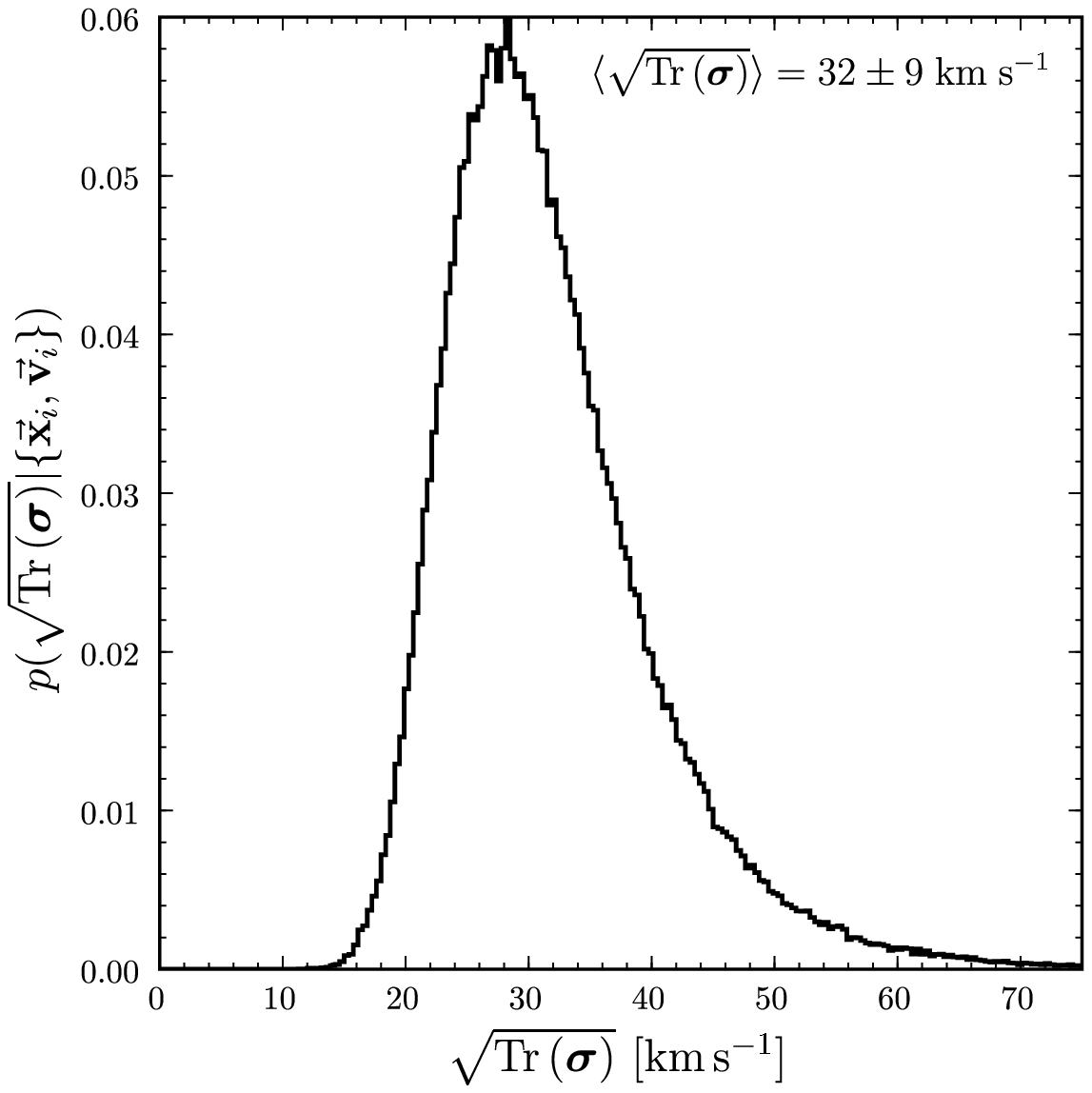}
\caption{Marginalized posterior probability distribution for the
parameters of the conditional velocity distribution of masers from
10$^6$ samples: mean motion toward the Galactic Center (\emph{top
left panel}); in the direction of Galactic rotation (\emph{top right
panel}); toward the North Galactic Pole (\emph{bottom left panel});
the square root of the trace of the velocity dispersion tensor
(\emph{bottom right panel}).}\label{fig:dist}
\end{figure}

At the same time, the MCMC procedure provides fully marginalized
posterior distributions for the parameters of the conditional velocity
distribution function of masers, which are given in
\figurename~\ref{fig:dist}: shown are the posterior distributions for
the three components of the mean offset from circular velocity of the
masers, \ie, the mean peculiar velocity, in cylindrical coordinates
(toward the Galactic Center, in the direction of Galactic rotation,
and toward the North Galactic Pole) as well as for the trace of the
velocity dispersion tensor. From this we confirm the mean lag of 15 km
s$^{-1}$---we find a lag of $14 \pm 5$ km s$^{-1}$---of the masers
with respect to their local standards of rest previously found by
\reid. \figurename~\ref{fig:dist} shows that the masers have a mean
velocity toward the Galactic Center of $7 \pm 6$ km s$^{-1}$. Taken
together, these mean peculiar velocities imply that the masers are
typically just past the apocenter of their orbits. We also find a mean
velocity component of $3 \pm 3$ km s$^{-1}$ in the direction toward
the North Galactic Pole.

From the posterior distribution for the trace of the velocity
dispersion tensor we see that the masers have a relative large
velocity dispersion---$\trace(\maserdisp)\sim\![29$ km
s$^{-1}$]$^2$---larger than might be expected from a comparison with
the velocity dispersion of young stars in the Solar neighborhood,
whose trace is about [14 km s$^{-1}$]$^2$ \citep{Hogg05a}. Since we
put no restrictions on the form of $\maserdisp$ we also obtain
posterior probability distributions for all of the components of
$\maserdisp$: for the diagonal components we find
$\sqrt{\maserdisp_{RR}} = 22 \pm 8$ km s$^{-1}$,
$\sqrt{\maserdisp_{\phi\phi}} = 18 \pm 7$ km s$^{-1}$, and
$\sqrt{\maserdisp_{zz}} = 12 \pm 5$ km s$^{-1}$. As we discuss below,
the fact that we obtain these large values could be because our model
for the conditional velocity distribution is too restrictive.

In order to assess the possible affect of outliers on our inference,
we have performed the same analysis assuming a distribution of the
peculiar velocities which consists of a mixture of two Gaussian
distributions, identical in every aspect except that the second
Gaussian has half of the weight and twice the dispersion of the first
Gaussian (by doubling each component of the velocity dispersion
tensor). We find the same posterior distributions for the dynamical
parameters and the mean offset; the inferred dispersion of the masers
is, predictably, somewhat smaller: the trace of the covariance matrix
peaks at [22 km s$^{-1}$]$^2$. Two specific candidate outliers, the
sources NGC 7538 and G 23.6-0.1, were identified and removed from the
sample by \reid, because they displayed large post-fit residuals. To
assess whether these two sources affect our results significantly, the
same analysis as described above of the \reid\ basic sample of 16
masers was performed, leaving out the sources NGC 7538 and
G~23.6-0.1. We find basically the same result: $\vc = 245 \pm 13$ km
s$^{-1}$. Thus, as opposed to \reid, who found that these two sources
significantly raise the circular velocity derived from the maser data,
our result is robust with respect to their inclusion.

\section{Discussion}\label{sec:discussion}

We have re-analyzed the recent maser kinematics from \reid, to see
what they tell us about $\vc(R_0)$ and the maser orbits. Our analysis
differs from that of \reid\ by allowing for a more general model for
the distribution of the velocities of the masers with respect to their
local standards of rest, by using a proper probabilistic framework
that includes proper marginalization over uninteresting parameters,
and by the explicit inclusion of suitable prior information. From
this, we find an estimate of $\vc$ of $244 \pm 13$ km s$^{-1}$, the
same value as the mode of our prior, and substantially lower than the
estimate of \reid. Our analysis has also shown that the current maser
measurements have only limited power to constrain $\vc$ beyond the
prior; dropping the prior coming from the measured proper motion of
Sgr A$^*$ we find $\vc = 246 \pm 30$ km s$^{-1}$; further dropping the
prior information on $R_0$, the maser data provide no constraint on
$\vc$ at all.

The value for $\vc$ that we have inferred in this paper from the
kinematics of Galactic masers compares favorably with other recent
measurements of the circular velocity. As is clear from
\figurename~\ref{fig:thetao}, the posterior probability distribution
for the circular velocity is peaked at about the same value as the
prior probability distribution obtained from combining the precise
measurements of the distance to the Galactic Center, the proper motion
of Sgr A$^*$, and the Solar motion in the direction of Galactic
rotation. It is also consistent with the value of $\vc = 221 \pm 18$
km s$^{-1}$ from a recent measurement based on the completely
different principle of fitting an orbit to the GD-1 stellar stream
\citep{Koposov09a}. Combining these estimates by inverse variance
weighting we find a value for the circular velocity of $\vc = 236 \pm
11$ km s$^{-1}$.

The results in this paper are unaffected by the uncertainty in the
value of the Solar motion with respect to the \lsr. If we use a larger
uncertainty in the value of $\vsunlsr$ of 3 km s$^{-1}$ in each
component, as suggested by an analysis of the effect of moving groups
on $\vsunlsr$ (J.~Bovy \& D.~W.~Hogg, in preparation), we retrieve the
same estimate $\vc = 244 \pm 14$ km $^{-1}$ as before. Even when we
use an uncertainty of 15 km s$^{-1}$ in the value of each component of
$\vsunlsr$, we find a slight increase in the uncertainty, but still
the same value $\vc = 244 \pm 20$ km s$^{-1}$. Thus, the uncertainty
in $\vsunlsr$ only affects our conclusions if it is larger than about
10 km s$^{-1}$.

\begin{figure}
\includegraphics[width=.45\textwidth]{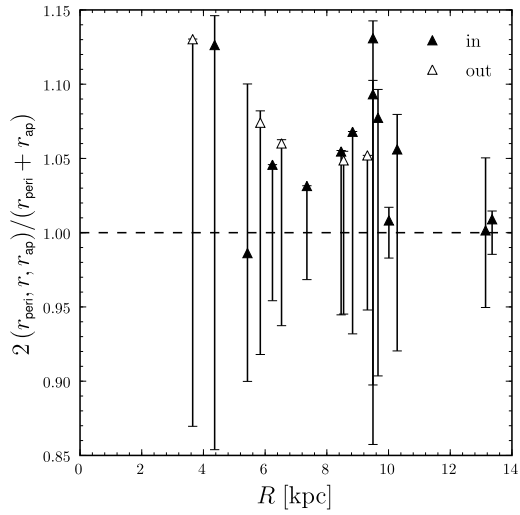}
\caption{Orbital eccentricities and phases of the observed masers in a
logarithmic potential: pericenter radius $r_{\mathrm{peri}}$,
apocenter radius $r_{\mathrm{ap}}$, and current radius of the masers,
normalized by the mean of the pericenter and apocenter radii, as a
function of Galactocentric radius in a spherically symmetric
logarithmic potential for $R_0 = 8.2$ kpc and $\vc = 244$ km
s$^{-1}$. Filled symbols indicate that the maser is moving toward the
Galactic Center.}\label{fig:phases}
\end{figure}

We also learned that the masers on average lag $\vc$ and are moving
toward the Galactic Center. This fact is illustrated in \figurename
s~\ref{fig:dist} and \ref{fig:phases}, where the orbital phases of the
masers are shown for a logarithmic potential $\Phi = \vc^2 \,\ln r$
\citep[\eg, \eqnname~(3.14) in][]{binneytremaine} assuming $R_0 = 8.2$
kpc and $\vc = 244$ km s$^{-1}$. This will be interesting to analyze
in the context of spiral shock models.

Our analysis implies that the present maser data do not lead to a
substantive improvement of our knowledge of \Ro\ and $\vc$, as most of
the information in the data is spent on determining the properties of
the conditional velocity distribution of the masers. It is also
remarkable that, given all of the prior information, the masers are
much more informative about \Ro\ than they are about the angular
rotation speed at the Sun's radius, as the posterior distribution for
$\Omega_0$ is barely distinguishable from the prior distribution.

Despite the fact that most of the information content in the maser
data is already being used to infer the distribution function, it is
possible that our model for the distribution function is not general
enough. For one, it is very likely that the distribution function of
the masers depends on the Galactocentric radius and, in particular,
that the mean velocity offset in the direction toward the Galactic
center depends on radius. Indeed, there is some indication of that
already in our results, as the large velocity dispersion of the masers
is mostly driven by a large velocity dispersion in the direction
toward the Galactic Center; this could be due to an unmodeled radial
dependence of the distribution function.

The measurement of the dynamics of the Galaxy performed here uses a
tracer population that is obviously non-angle mixed but has no
unambiguous non-angle-mixed interpretation---such as a stellar stream
tracing out an orbit. Such a measurement has the fundamental problem
that structure in the distribution function of the tracers is, in a
sense, exchangeable with complexity of the potential. Therefore,
detailed measurements of the potential of the Galaxy using larger
samples of masers will very likely be fundamentally limited by our
lack of knowledge about the distribution function of the masers. As
more masers with precise kinematic information become available---as
many as 400 are possible over the next few years (M.~Reid, private
communication)---more detailed inferences of the distribution function
will have to be made simultaneously with more precise measurements of
the potential of the Galaxy from these masers. The method described
and used in this paper is flexible enough to handle these more general
distribution functions and more general models for the potential of
the Galaxy.

\acknowledgments It is a pleasure to thank Dustin Lang, Mark Reid, and
Scott Tremaine for helpful discussions and the anonymous referee for
valuable comments. JB and DWH were partially supported by NASA (grant
NNX08AJ48G). JB was partially supported by New York University's
Horizon fellowship. DWH is a research fellow of the Alexander von
Humboldt Foundation of Germany. JB was partially supported by the
Max-Planck-Institut f\"ur Astronomie, and is grateful for its
hospitality during part of the period during which this research was
performed. JB also gratefully acknowledges the hospitality of the
Lorentz Center (Leiden) where parts of this research were performed.

\appendix

\section{MCMC exploration of the posterior distribution}\label{sec:mcmc}

We explore the posterior probability distribution using a
Metropolis-Hastings (MH) MCMC algorithm \citep[\eg,][]{mackay}. The MH
algorithm works by proposing new model parameters $x'$ from a proposal
distribution $Q(x';x^{(t)})$ that can only depend on the current
values $x^{(t)}$ of the parameters. One then computes the quantity
\begin{equation}
a =
\frac{p(x'|\setofxv)\,Q(x^{(t)};x')}{p(x^{(t)}|\setofxv)\,Q(x';x^{(t)})}\,.
\end{equation}
If $a \geq 1$ one accepts the new state; if $a < 1$, the new state is
accepted with probability $a$. If the new state is rejected, the old
state is added again as a sample of the posterior. This procedure
converges to give samples from the posterior.

As proposal distributions we use: (1) the prior for the components of
\vsunlsr, (2) a Gaussian for \Ro\ and $\pmsgra$ centered on the
current values with widths of 0.5 kpc and 0.12 km s$^{-1}$ kpc$^{-1}$,
respectively, (3) a Gaussian for the mean offset centered on the
current values with a width of $\sim\!10$ km s$^{-1}$ for each
component, and (4) a Wishart distribution for the velocity dispersion
tensor with mean equal to the current tensor and shape parameter
$\sim\!20$. The widths of these last three proposal distributions were
chosen so as to give an acceptable acceptance rate of about
50\,percent. Monte Carlo chains were run with different sets of
parameters of the proposal distributions and the resulting posterior
probability distributions were found to be independent of the
parameters of the proposal distributions.

\clearpage

\clearpage

\clearpage

\clearpage

\end{document}